\documentclass[preprint,aps,12pt,preprintnumbers,eqsecnum,nofootinbib,superscriptaddress]{revtex4}
\usepackage{tabularx} 
\usepackage{amsmath}  
\usepackage{graphicx} 
\usepackage[margin=1in,letterpaper]{geometry} 
\usepackage{physics}
\usepackage{ amssymb }
\usepackage[final]{hyperref} 
\usepackage{simplewick}
\hypersetup{
	colorlinks=true,       
	linkcolor=blue,        
	citecolor=blue,        
	filecolor=magenta,     
	urlcolor=blue         
}
\begin{document}

\title{\vspace*{0.5in} 
Background-Independent Composite Gravity
\vskip 0.1in}
\author{Austin Batz}\email[]{ajbatz@email.wm.edu}\affiliation{High Energy Theory Group, Department of Physics,
William \& Mary, Williamsburg, VA 23187-8795, USA}

\author{Joshua Erlich}\email[]{erlich@physics.wm.edu}\affiliation{High Energy Theory Group, Department of Physics,
William \& Mary, Williamsburg, VA 23187-8795, USA}

\author{Luke Mrini}\email[]{lamrini@email.wm.edu}\affiliation{High Energy Theory Group, Department of Physics,
William \& Mary, Williamsburg, VA 23187-8795, USA}

\date{November 12, 2020}


\begin{abstract}
We explore a background-independent theory of composite gravity. The vacuum expectation value of the composite metric satisfies Einstein's equations (with corrections) as a consistency condition, and selects the vacuum spacetime. A gravitational interaction then emerges in vacuum correlation functions. The action remains diffeomorphism invariant even as perturbation theory is organized about the dynamically selected 
 vacuum spacetime.
We discuss the role of nondynamical clock and rod fields in the analysis, the identification of physical observables, and the generalization to other theories including the standard model. \end{abstract}
\maketitle

\section{Introduction}

\vspace{-0.5\baselineskip}
Diffeomorphism invariance is responsible for a number of  conceptual and technical challenges for quantum gravity. The Hamiltonian in diffeomorphism-invariant theories vanishes in the absence of a spacetime boundary, and is replaced by a constraint. In canonical quantum gravity \cite{DeWitt:1967yk}, the functional Schr\"odinger equation is replaced by the Wheeler-DeWitt equation, which imposes the Hamiltonian constraint on states. The absence of an {\em a priori} notion of time evolution in such a description suggests a relational approach to dynamics in which a clock is identified from within the system under investigation \cite{Page:1983uc}. The challenge of identifying a  clock with which to describe dynamics in quantum gravity is  the well-known problem of time \cite{Isham:1992ms}.

Also related to diffeomorphism invariance is the puzzle of identifying observables. It is current wisdom that there are no local observables in a diffeomorphism-invariant theory. Even scalars under diffeomorphisms, such as the curvature scalar, have a different profile if spacetime points are dragged 
via a diffeomorphism. Integrals of suitable local operators \cite{Giddings:2005id}, or appropriately dressed operators \cite{Donnelly:2015hta}, could serve as diffeomorphism-invariant observables as long as contact can be made with the semiclassical world of our observations and experiments. The concept of an observable is a semiclassical construct, and in a semiclassical context there is less difficulty in defining local observables in a specified vacuum. Indeed, in a composite gravity scenario the identification of the spacetime metric as a composite operator makes possible well-defined correlation functions of local operators.

Perturbative approaches to quantum gravity, such as perturbative string theory, concern fluctuations about a  background spacetime that appears explicitly in the description of the dynamics, for example in the action functional. Nonperturbative aspects of string theory that are well understood, including  the AdS/CFT correspondence, 
also contain a background spacetime in their description. In this sense, string theory in its present state is not background-independent.  Einstein's equations (plus corrections) arise as a self-consistency condition on the background spacetime, namely conformal invariance of the worldsheet field theory. Analogously, despite the absence of conformal symmetry in the class of theories considered in this paper, cancellation of tadpoles will require that the vacuum spacetime satisfy Einstein's equations up to corrections. Also as in string theory, a gravitational interaction emerges dynamically in a composite gravity scenario from fluctuations of the fundamental degrees of freedom.

In composite gravity, the spacetime metric is identified as a composite operator that depends on the elementary fields and  their derivatives. The spacetime geometry is not integrated over separately in the functional integral, but rather each configuration of the fundamental fields corresponds to a unique metric at the classical level, one with no obvious relation to Einstein's equations.  In such a scenario, specifying the form of the vielbein or metric in vacuum requires consideration of both short and long-distance physics. 
Both the vacuum expectation value of the composite metric and
correlation functions involving the composite metric require regularization due to the  products of fields at the same point that define the composite metric.  The vacuum expectation value of the composite metric is dominated by ultraviolet regulator-scale physics but is determined everywhere in spacetime. Hence, macroscopic diffeomorphisms act nontrivially on the short-distance contribution to expectation values and correlation functions.  Diffeomorphisms that transform the fundamental fields also transform the composite metric operator and its expectation value. The gauge-fixing of diffeomorphisms is implemented not in the selection of field configurations that contribute to the functional integral, but rather in the identification of the spacetime metric as a function of the fundamental fields, and in the process of regularization when specifying the vacuum metric. The vacuum provides a semiclassical description of spacetime, addressing the problem of time. Formally, we will introduce {\em nondynamical} clock and rod fields to fix a coordinate basis in the vacuum spacetime. Local observables are well defined in this semiclassical arena.

Perturbation theory in the vacuum then proceeds via curved-space quantum field theory methods, by way of which we will demonstrate the existence of an emergent long-range gravitational interaction. We presume the existence of a physical ultraviolet regulator that would complete this  framework, but in the present work as a proxy for a physical regulator we use dimensional regularization in the DeWitt-Schwinger representation of Green's functions. 
 The search for a quantum theory of gravity is replaced by the search for a physical diffeomorphism-invariant regulator; there is no need to begin with the Einstein-Hilbert action on the road to quantum gravity. We will briefly discuss possibilities for new short-distance physics that could play the role of a physical regulator and complete this framework.

Over the years there have been a number of investigations of a diffeomorphism-invariant scalar toy model described by the action,
\begin{equation}
S=\int d^Dx \ \bigg(\frac{\frac{D}{2}-1}{V(\phi^a)}\bigg)^{\frac{D}{2}-1}\sqrt{\left|\det\bigg(\sum_{a=1}^N \partial_\mu\phi^a\partial_\nu\phi^a\bigg)\right|}, \label{eq:S}
\end{equation}
where  we consider potentials $V(\phi^a)$ of the form, \begin{equation}
V(\phi^a)=V_0'+ \sum_{a=1}^N\frac12m^2\phi^a \phi^a. \label{eq:Vphi}\end{equation}
We write the dimensionful parameter $V_0'$ as the sum of three terms:
\begin{equation}
    V_0' \equiv \; V_0+\Delta V_{{\rm ct}}+\Lambda.
\end{equation}
The counterterm $\Delta V_{\rm ct}$ will be chosen such that in the vacuum $\Delta V_{\rm ct}+\langle \tfrac12 m^2 \phi^2\rangle=0$. The constants $V_0$ and $\Lambda$  will be defined in the following section.

 This action is reminiscent of the scalar part of the Dirac-Born-Infeld action describing D-brane dynamics.
If there is a physical ultraviolet regulator for the quantum field theory with this action, then with a certain tuning of the constant part of $V(\phi^a)$  the theory has been demonstrated to include an emergent long-range gravitational interaction in Minkowski space. The emergent gravitational interaction was anticipated \cite{Akama:1978pg,Akama:2013tua} by comparison with Sakharov's induced gravity \cite{Sakharov:1967pk}. In order to demonstrate the emergent gravitational interaction, the analysis in Ref.~\cite{Carone:2016tup} adds scalar clock and rod fields to the theory, which allowed for a perturbative expansion about a Minkowski-space vacuum after a tuning of parameters.
A similar analysis demonstrated an emergent gravitational interaction in a theory with scalars and fermions reminiscent of the supersymmetric D-brane action \cite{Carone:2018ynf}, and in a curved background perturbed about Minkowski space \cite{Chaurasia:2017ufl}.
Ref.~\cite{Carone:2017mdw} explored nonlinear gravitational self-interactions in the same theory, and also pointed out an interesting fact: a renormalization that was introduced in order to cancel tadpole diagrams also rescaled the clock and rod field configuration to zero. In hindsight the reason for this is clear: a spacetime-dependent background for fields explicitly breaks diffeomorphism invariance. The clock and rod fields serve as a  tool for providing a description of the vacuum about which to consider fluctuations, but ultimately they appear in the action only in terms which identically cancel one another, and do not participate in the dynamics \cite{Carone:2019xot}.

In this paper, we probe the theory by detuning the vacuum energy  in order to understand the selection of a self-consistent vacuum and the emergence of a long-range gravitational interaction in this more-general setting. We use the adiabatic expansion of curved-space Green's functions in the DeWitt-Schwinger proper-time representation in order to analyze ultraviolet divergences in the correlation functions  which, when regularized by dimensional regularization as a proxy for a physical regulator,  are responsible for an emergent long-range gravitational interaction. Because the vacuum provides a spacetime background in which we can calculate correlation functions, diffeomorphism invariance does not lead to difficulty in the identification of observables.

In Section~\ref{sec:Model} we describe the toy model of composite gravity and calculate the vacuum expectation value of the composite metric. In Section~\ref{sec:Emergent} we derive the emergent gravitational interaction in correlation functions. In Section~\ref{sec:Discussion} we discuss these results and comment on quantum gravity corrections as $1/N$ corrections. We conclude in Section~\ref{sec:Conclusions} with some comments on implications of composite gravity for cosmology of the early universe.

\section{The Model}
\label{sec:Model}
Similar to the relation between the Polyakov and Nambu-Goto formulations of the bosonic string and as in the construction of Refs.~\cite{Akama:1978pg,Akama:2013tua,Carone:2016tup},  we note that that the action Eq.~(\ref{eq:S}) is equivalent to the action
\begin{equation} \label{eq:action1p}
    S = \int \dd^D x \sqrt{\abs{g}} \quantity[\frac{1}{2}g^{\mu \nu}\bigg(\sum_{a=1}^N \partial_{\mu} \phi^a \partial_{\nu} \phi^a \bigg)-V(\phi^a)],
\end{equation}
with the composite metric identified as, 
\begin{eqnarray}
    g_{\mu\nu}&=&\frac{D/2-1}{V(\phi^a)}\left(\sum_{a=1}^N\partial_\mu\phi^a\partial_\nu\phi^a\right) \nonumber \\
    &=&\frac{D/2-1}{V(\phi^a)}\left(\sum_{M,N=1}^D \partial_\mu X^M\partial_\nu X^N G_{MN}(\{X\})+\left[\sum_{a=1}^N\partial_\mu\phi^a\partial_\nu\phi^a\right.\right.  \nonumber\\
    &&-\left.\left.\sum_{M,N=1}^D \partial_\mu X^M\partial_\nu X^N G_{MN}(\{X\})\right]\right). \label{eq:gmn}
\end{eqnarray}

We have introduced the clock and rod fields $X^M$, which appear in two terms that identically cancel one another in Eq.~(\ref{eq:gmn}), generalizing a related procedure in Ref.~\cite{Carone:2019xot}.  The clock and rod fields are
nondynamical, but serve to define a coordinate basis for the vacuum spacetime.
We choose $G_{MN}$ and $X^M$ such that the vacuum spacetime metric in a coordinate basis specified by $X^M(x)$ takes the form, \begin{equation}\label{eq:Gmn}
G_{\mu\nu}(x)\equiv\frac{D/2-1}{V_0+\Lambda}\sum_{M,N=1}^D \partial_\mu X^M\partial_\nu X^N G_{MN}[\{X(x)\}]. \end{equation}
More precisely, we will choose $G_{\mu\nu}(x)$ self-consistently as the regularized vacuum expectation value of the composite metric $\langle g_{\mu\nu}\rangle$ in the coordinate basis determined by the clock and rod fields $X^M(x)$.
The vacuum metric $G_{\mu\nu}(x)$ transforms covariantly under coordinate transformations of the clock and rod fields.
In the static-gauge basis normalized as, \begin{equation}
\partial_\mu X^M=\sqrt{\frac{V_0+\Lambda}{D/2-1}} \delta_\mu^{\ M}, \end{equation}
$G_{MN}$ is chosen so that the regularized vacuum expectation value of the composite metric is $\langle g_{\mu\nu}\rangle=G_{MN}\delta_\mu^{\ M}\delta_\nu^{\ N}$. 
The terms in brackets in Eq.~(\ref{eq:gmn}) will collectively represent a perturbation
about the first term, as the would-be divergent vacuum expectation values of the terms in brackets will cancel in the self-consistent vacuum spacetime.

We will be interested in the ultraviolet divergences in  vacuum expectation values.
We calculate the regularized divergences  by way of an adiabatic expansion in the DeWitt-Schwinger representation of curved-space Green's functions. To perturb about the vacuum, we introduce the vacuum spacetime into the action Eq.~(\ref{eq:S})  by analogy with Eq.~(\ref{eq:gmn}):
\begin{eqnarray}
S=\int d^Dx \ &&\left(\frac{\frac{D}{2}-1}{V(\phi^a)}\right)^{\frac{D}{2}-1}\left| \det\left( \sum_{M,N=1}^D \partial_\mu X^M\partial_\nu X^N G_{MN}(\{X\}) 
\right.\right. \nonumber \\ 
&&\left.\left.+ \left[\sum_{a=1}^N \partial_\mu\phi^a\partial_\nu\phi^a - \sum_{M,N=1}^D \partial_\mu X^M\partial_\nu X^N G_{MN}(\{X\})\right] \right)
\right|^{1/2}. \label{eq:S2}
\end{eqnarray}
In the functional integral we can fix a coordinate basis for the vacuum spacetime by substituting for the clock and rod fields a profile $X^M(x)$, or equivalently by integrating over the {\em nondynamical} clock and rod fields $X^M(x)$ and inserting a gauge-fixing delta function with trivial Fadeev-Popov determinant. For example, we write the partition function as,
\begin{equation}
Z=\int {\cal D} X^M(x) \int {\cal D}\phi^a(x) \ e^{iS}\,\delta\left(X^M-\sqrt{\frac{V_0+\Lambda}{D/2-1}}\,\delta_\mu^M\,x^\mu\right). \label{eq:ZX}
\end{equation}
The integral over $X^M$ is trivial, as the action is independent of $X^M$. However, this manipulation facilitates a perturbative expansion in which the vacuum spacetime metric is specified in a fixed coordinate basis. The profile for
the clock and rod fields in Eq.~(\ref{eq:ZX}) sets the vacuum expectation value of the composite metric $g_{\mu\nu}$ to
$G_{MN}\,\delta_\mu^{\ M}\delta_\nu^{\ N}$. Any configuration of the clock and rod fields without critical points where $\det(\partial_\mu X^M)=0$ can be transformed to the configuration fixed by the delta function in Eq.~(\ref{eq:ZX}) by an invertible coordinate transformation.

We are aided by previous calculation of the regularized effective action for free fields in curved spacetime backgrounds. 
The effective action $W_{\rm eff}$ defined by, \begin{equation}
\int {\cal D}\phi^a\,e^{iS_{\rm free}[g_{\mu\nu},\phi^a]}=e^{iW_{\rm eff}[g_{\mu\nu}]}, \end{equation}
where \begin{equation}
S_{\rm free}=\int d^Dx\,\sqrt{|g|} \left(\frac12 \sum_{a=1}^N\, \partial_\mu\phi^a\partial_\nu\phi^a\,g^{\mu\nu}-\frac{m^2}{2}\sum_{a=1}^N\phi^a\phi^a\right),
\end{equation}
determines connected correlation functions of products of the curved-space free-field energy-momentum tensor, {which in the background specified by $G_{\mu\nu}$ takes the form,}
\begin{equation} \label{eq:Tmn}
    \mathcal{T}_{\mu \nu} = \sum_{a=1}^N \bigg( \partial_{\mu}\phi^a\partial_{\nu}\phi^a - \frac{1}{2}G_{\mu \nu}\bigg(\partial^{\alpha}\phi^a\partial_{\alpha}\phi^a - m^2\phi^a \phi^a \bigg)\bigg). 
\end{equation}

Tadpoles attached to Feynman diagrams contributing to correlation functions arise from either $\langle \phi^2\rangle\equiv iG_F(x,x)
$
or $\langle\partial_\mu\phi(x)\partial_\nu\phi(x)\rangle
$.
The scalar tadpole $iG_F(x,x)$
can be regularized by dimensional regularization by keeping $D\neq4$ and extracting poles in $D-4$. 
The divergences depend on the curvature tensor and its derivatives.

The Green's function satisfies the equation,
\begin{equation} \label{eq:eom}
    (\square+m^2)G_F(x,x')=-\frac{1}{\sqrt{\abs{g(x)}}}\delta^{(D)}(x-x'),
\end{equation}
where $\square$ is the curved-space d'Alembertian, $\square\equiv g^{\mu\nu}\nabla_\mu\nabla_\nu$. In the DeWitt-Schwinger proper time representation, the Green's function takes the form \cite{DeWitt:1975ys},
\begin{equation} \label{eq:ds}
    G_F(x,x') = \int_0^\infty \dd s \: G^{\text{DS}}_F(x,x';is),
\end{equation}
with an adiabatic expansion (an expansion in derivatives of the background metric) given by
\begin{equation} \label{eq:adiabatic}
    G^{\text{DS}}_F(x,x';is) = \frac{\sqrt{\Delta(x,x')}}{(4\pi)^{D/2}}\frac{1}{(is)^{D/2}}e^{-im^2s+\sigma/(2is)}\sum_{j=0}^\infty a_j(x,x')(is)^j
\end{equation}
where $\sigma(x,x')$ is the geodetic squared distance between $x$ and $x'$ in the background spacetime,
and $\Delta(x,x')$ is the Van Vleck determinant
\begin{equation} \label{eq:vanvleck}
    \Delta(x,x') = -\frac{\text{det}(\partial_\mu \partial_\nu \sigma(x,x'))}{\sqrt{g(x)g(x')}}.
\end{equation}
The Seeley-DeWitt coefficients $a_j$ can be solved for recursively using Eqn.~(\ref{eq:eom}) with $a_0(x,x')=1$.
We will ultimately be concerned with the limit in which both arguments are evaluated at the same point. The first three $a_j(x)\equiv a_j(x,x)$, which depend on the curvature tensor and its derivatives,  have the values \cite{Christensen:1976vb}
\begin{eqnarray} 
a_0(x)&=&1 \nonumber \\
a_1(x)&=&\tfrac16 R \label{eq:ak} \\
a_2(x)&=&\tfrac{1}{180}R^{\mu\nu\rho\sigma}R_{\mu\nu\rho\sigma}-\tfrac{1}{180}R^{\mu\nu}R_{\mu\nu}+\tfrac{1}{30}\Box R+\tfrac{1}{72}R^2. \nonumber
\end{eqnarray}

The effective action for the free theory in the curved-space vacuum was studied in the adiabatic expansion by Christensen \cite{Christensen:1978yd}, Bunch \cite{Bunch:1979mq}, and others. The calculation of the effective action proceeds by identifying the adiabatic coefficients in an expansion based on the DeWitt-Schwinger proper time representation \cite{DeWitt:1975ys},\begin{equation}
W_{\rm eff}=\int \frac{d^Dx}{2(4\pi)^{D/2}} \sqrt{g}\,(is)^{1+D/2} e^{-im^2 s}F(x,is), \label{eq:Weff1}
\end{equation}
with \begin{equation}
F(x,is)=\sum_{k=0}^{\infty}a_k(x)(is)^k.
\end{equation}
The coefficients $a_k(x)$ were calculated in Ref.~\cite{Christensen:1978yd}, given by Eq.~(\ref{eq:ak}).
Integrating Eq.~(\ref{eq:Weff1}) over $s$, the  adiabatic expansion of the effective action takes the form  \cite{Bunch:1979mq}:
\begin{equation}
W_{\rm eff}[g_{\mu\nu}]=N\int d^D x\sqrt{|g|}\left[\frac{1}{2(4\pi)^{D/2}}\sum_{k=0}^{\infty}a_k(x)m^{D-2k}\Gamma(k-D/2)
\right], \label{eq:Weff}
\end{equation}

The three coefficients $a_0$, $a_1$ and $a_2$ multiply divergences from
$\Gamma(k-D/2)$ as $d\rightarrow 4$, while the remaining terms in the expansion of the effective action are finite in the same limit. Expanding in powers of $(D-4)$, the poles in the effective action take the form  \cite{Bunch:1979mq}:
\begin{eqnarray}W_{\rm div}&=&N\int d^Dx\sqrt{|g|}\left\{-\frac{1}{(4\pi)^{D/2}}\,\frac{1}{D-4}
\left[\frac{4m^D\, a_0}{D(D-2)}-\frac{2m^{D-2}\,a_1}{D-2}+a_2\right]\right\} \nonumber \\
&=&N\int d^Dx\sqrt{|g|}\left\{-\frac{1}{(4\pi)^{D/2}}\,\frac{1}{D-4}
\left[\frac{4m^D}{D(D-2)}-\frac{m^{D-2}\,R}{3(D-2)}+a_2\right]\right\}  \label{eq:Wdiv}
\end{eqnarray}
The pole terms in $\langle 
\mathcal{T}_{\mu\nu}\rangle\equiv \,  _{\rm out}\hspace{-1pt}\langle0|\mathcal{T}_{\mu\nu}|0\rangle_{\rm in}/_{\rm out}\langle0|0\rangle_{\rm in}$ are then,
\begin{eqnarray}
\langle 
\mathcal{T}_{\mu\nu} \rangle&=&\frac{2}{\sqrt{|g|}}\frac{\delta W_{\rm div}^{(2)}}{\delta g^{\mu\nu}(x)}\Bigg|_{g_{\alpha\beta}=G_{\alpha\beta}} \\
&=&\frac{1}{(4\pi)^{D/2}}\frac{N}{D-4}\left(\frac{4m^{D-2}}{D(D-2)}\right)\left[m^2 G_{\mu\nu}+\frac{D}{6}\left(R_{\mu\nu}-\tfrac12 G_{\mu\nu}R\right)\right]+\Delta\langle\mathcal{T}_{\mu\nu}\rangle, \label{eq:<Tmn>}
\end{eqnarray}
where $\Delta\langle\mathcal{T}_{\mu\nu}\rangle$ contains the four-derivative terms from variation of $a_2(x)$, which we record here for
completeness \cite{Bunch:1979mq}:
\begin{eqnarray}
\Delta\langle\mathcal{T}_{\mu\nu}\rangle=-\frac{1}{(4\pi)^{D/2}}\,\frac{1}{D-4}\left[\frac{1}{90}H_{\mu\nu}+\frac{1}{36}\,^{(1)}H_{\mu\nu}
-\frac{1}{90} \,^{(2)}H_{\mu\nu}\right], \end{eqnarray}
where, \begin{eqnarray}
H_{\mu\nu}&=&\frac{1}{2}G_{\mu\nu}R^{\alpha\beta\gamma\delta}R_{\alpha\beta\gamma\delta}-2R_{\mu\alpha\beta\gamma}R_{\nu}^{\ \alpha\beta\gamma}-4\square R_{\mu\nu}+2R_{;\mu\nu}+4R_{\mu\alpha}R_{\nu}^{\ \alpha}-4R^{\alpha\beta}R_{\alpha\mu\beta\nu}, \nonumber \\
^{(1)}H_{\mu\nu}&=&2R_{;\mu\nu}-2G_{\mu\nu}\square R+\frac12 R^2 G_{\mu\nu}-2RR_{\mu\nu}, \nonumber \\
^{(2)}H_{\mu\nu}&=&\frac{1}{2}R^{\alpha\beta}R_{\alpha\beta}\,
G_{\mu\nu}-2R^{\alpha\beta}R_{\alpha\mu\beta\nu}+R_{;\mu\nu}-\square R_{\mu\nu}-\frac12G_{\mu\nu}\square R .
\end{eqnarray}

We will be most concerned with the terms up to adiabatic order 2, {\em i.e.} the terms arising from $a_0$ and $a_1$, and from now on we will drop terms arising from $a_2$. The corrections are interesting from a cosmological perspective, but our immediate interest is in the emergent gravitational interaction and dynamical selection of the vacuum, which are not sensitive to these corrections as long as the curvature of the vacuum spacetime is small compared to $m^2$. 

We can obtain the scalar tadpole directly or from the effective action using, \begin{equation}
\langle \phi^2\rangle=-\frac{2}{\sqrt{|g|}}\frac{\delta W_{\rm eff}}{\delta m^2},\end{equation}
where we treat $m^2$ as a source for the operator $(-\phi^2/2)$.
The terms to second order in the adiabatic expansion are,
\begin{equation}
\langle\phi^2\rangle=\frac{1}{(4\pi)^{D/2}}\frac{N}{D-4}\frac{4m^{D-4}}{(D-2)}\left[m^2-\frac{D-2}{12}R\right].
\end{equation}
Comparison with the trace of Eq.~(\ref{eq:<Tmn>}) demonstrates that, to second adiabatic order, \begin{equation}
\langle 
\mathcal{T}^\mu_{\ \mu}\rangle=m^2\langle\phi^2\rangle. \label{eq:Tm2phi2} \end{equation}
To calculate the regularized divergence in the remaining tadpole $\langle\partial_\mu\phi(x)\partial_\nu\phi(x)\rangle$, we use, \begin{equation}
\partial_\mu\phi\,\partial_\nu\phi=\mathcal{T}_{\mu\nu}-\frac{g_{\mu\nu}}{D-2}\left(\mathcal{T}^{\alpha}_{\ \alpha}-m^2\phi^2\right). \label{eq:dphidphi}\end{equation} 
From Eq.~(\ref{eq:Tm2phi2}) we then have, at the same order,
\begin{eqnarray}
\langle\partial_\mu\phi\,\partial_\nu\phi\rangle&=&\langle \mathcal{T}_{\mu\nu}\rangle \label{eq:<dphidphi>}  \nonumber \\ 
&=&\frac{1}{(4\pi)^{D/2}}\frac{N}{D-4}\left(\frac{4m^{D-2}}{D(D-2)}\right)\left[m^2 G_{\mu\nu}+\frac{D}{6}\left(R_{\mu\nu}-\tfrac12 G_{\mu\nu}R\right)\right].
\end{eqnarray}
The condition $\langle T_{\mu\nu}\rangle=\langle \mathcal{T}_{\mu\nu}\rangle+V_0'\, G_{\mu\nu}=0$, which includes the bare vacuum energy contribution from Eq.~(\ref{eq:action1p}), determines the self-consistent vacuum metric $G_{\mu\nu}$.
We note that in our organization of the perturbative expansion of correlation functions, from the form of the action Eq.~(\ref{eq:S2}), 
the $\langle\partial_\mu\phi\,\partial_\nu\phi\rangle$ tadpole always appears in the combination,
\[
\langle\partial_\mu\phi\partial_\nu\phi\rangle-\frac{2}{D-2}\left(V_0+\Lambda\right)G_{\mu\nu}.\]
Cancellation of the tadpole to second adiabatic order then requires: \begin{equation}
\frac{1}{(4\pi)^{D/2}}\frac{N}{D-4}\left(\frac{4m^{D-2}}{D(D-2)}\right)\left[m^2 G_{\mu\nu}+\frac{D}{6}\left(R_{\mu\nu}-\tfrac12 G_{\mu\nu}R\right)\right]-\frac{2}{D-2}\left(V_0+\Lambda\right)G_{\mu\nu}=0.
\end{equation}
Choosing \begin{equation}
V_0=\frac{1}{(4\pi)^{D/2}}\frac{N}{D-4}\,\frac{2}{D}m^D \end{equation}
cancels the term in the tadpole proportional to $m^D$. What remains of the tadpole cancellation condition is Einstein's equation with cosmological constant determined by $\Lambda$: \begin{equation}
\frac{1}{(4\pi)^{D/2}}\frac{N}{D-4}\left(\frac{m^{D-2}}{3}\right)\left(R_{\mu\nu}-\tfrac12 G_{\mu\nu}R\right)=\Lambda G_{\mu\nu}. \label{eq:Einstein}
\end{equation}
Eq.~(\ref{eq:Einstein}) receives four-derivative corrections from the Seeley-DeWitt coefficient $a_2$ that we dropped for simplicity of presentation, but it is straightforward to include those terms in the analysis.
We identify an effective Planck constant that we will see also determines the strength of the emergent gravitational interaction:\begin{equation}
\left(M_{\rm P}\right)^{D-2}=\frac{1}{(4\pi)^{D/2}}\frac{N}{D-4}\left(\frac{m^{D-2}}{3}\right)=\frac{V_0 D}{6m^2}.  \label{eq:MP} \end{equation}

As in Ref.~\cite{Carone:2016tup}, we also note that the tadpole  $\tfrac12m^2\langle\phi^2\rangle$ appears in combination with $\Delta V_{\rm ct}$ in the action Eq.~(\ref{eq:Vphi}). It follows from the trace of Eq.~(\ref{eq:Einstein}) that the curvature scalar is independent of $x$, and we can fix the counterterm \begin{equation}
\Delta V_{\rm ct}=-\tfrac12m^2\langle\phi^2\rangle=-\tfrac12\langle \mathcal{T}^\alpha_{\ \alpha}\rangle=-\frac{D}{D-2}(V_0+\Lambda).\end{equation}
In this case tadpoles are cancelled by the constant terms in the potential $V(\phi)$.

Under these circumstances, we also have $\langle \mathcal{T}_{\mu\nu}\rangle+(V_0+\Delta V_{ct}+\Lambda)G_{\mu\nu}=0$. In other words, cancellation of tadpoles in this theory is tantamount to vanishing of the expectation value of the full energy-momentum tensor including the vacuum energy contributions from Eq.~(\ref{eq:action1p}). This ensures that the short-distance physics does not generate a nonvanishing total energy-momentum tensor, which would break the diffeomorphism invariance of the classical theory. This is why Eq.~(\ref{eq:Einstein}) determines the self-consistent vacuum spacetime.

\section{The Emergent Gravitational Interaction}
\label{sec:Emergent}

We now consider the four-point correlation function of scalar fields in the self-consistent vacuum with spacetime metric $G_{\mu\nu}$ selected to enforce the condition $\langle T_{\mu\nu}\rangle=0$ via Eq.~(\ref{eq:Einstein}).
We expand the action Eq.~(\ref{eq:action1p}), writing the composite metric operator as $g_{\mu \nu}=G_{\mu \nu}+h_{\mu\nu}$. The action takes the form,
\begin{align} \label{eq:action3}
    S = \int & \dd^D x \sqrt{\abs{G}}{\bigg( 1+\frac{1}{2}h_{\alpha}^{\alpha} -\frac{1}{4}h_{\alpha \beta}h^{\alpha \beta} +\frac{1}{8}(h_{\alpha}^{\alpha})^2 + \dotsb \bigg)} \nonumber \\
    & \times \bigg[\frac{1}{2}\bigg( G^{\mu \nu}-h^{\mu \nu}+h^{\mu}_{\ \gamma}h^{\gamma \nu}+ \dotsb \bigg)\bigg(\sum_{a=1}^N \partial_{\mu} \phi^a \partial_{\nu} \phi^a 
    \bigg) - V_0' - \frac{1}{2}\sum_{a=1}^N m^2 \phi^a \phi^a\bigg],
\end{align}
where indices are raised and lowered by $G_{\mu \nu}$, and  normal ordering is meant to indicate that the expectation value of the operator has been subtracted off, {\em e.g.} $
    :\phi^2(x):\equiv \phi^2(x)-\langle \phi^2(x)\rangle
    $.

\begin{figure}[ht] 
        \includegraphics[width=1.0\columnwidth]{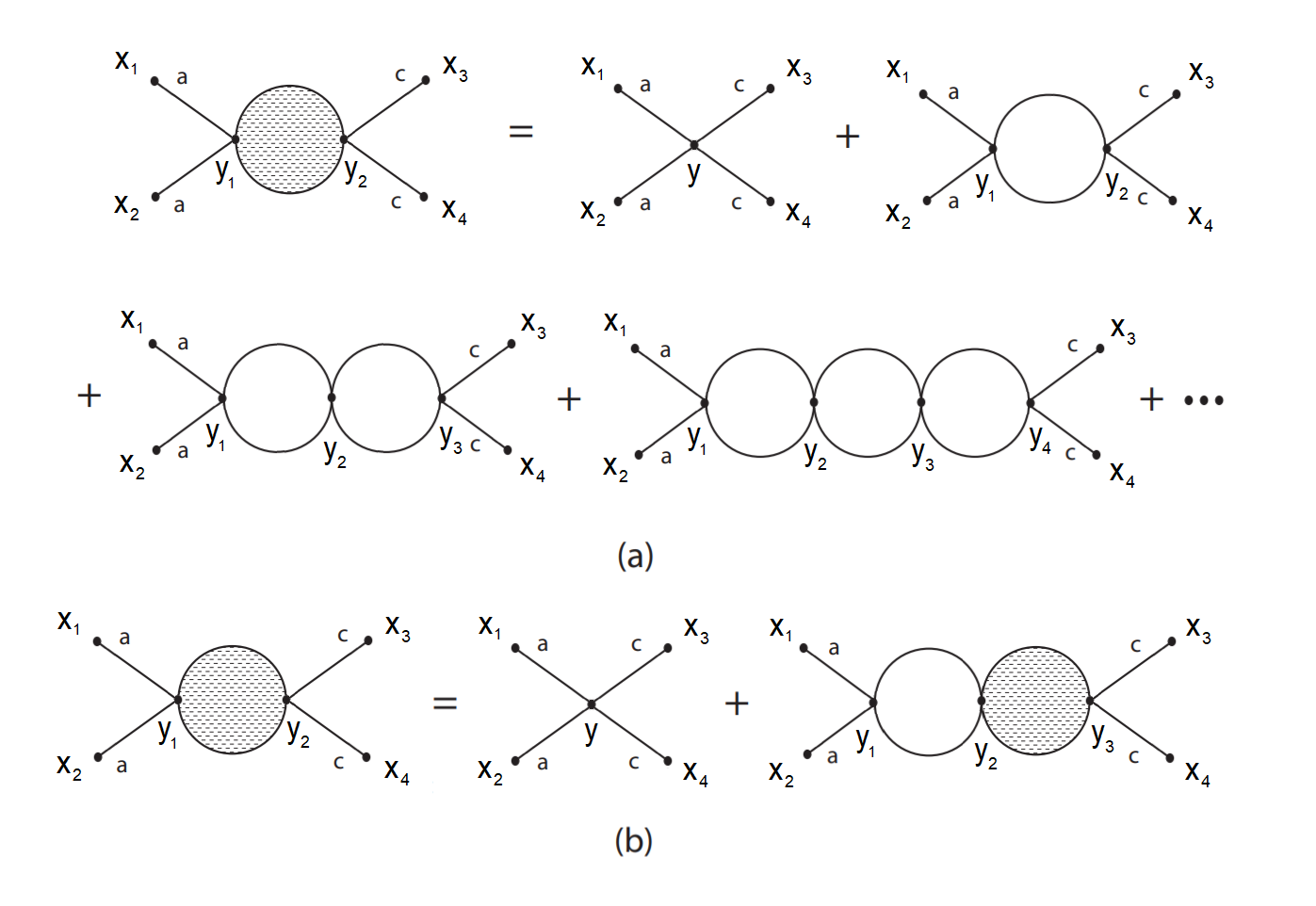}
        \caption{
                \label{fig:pos} 
                Contributions to the scalar 4-point function. (a) Diagrams that contribute at leading order in $1/N$. (b) The equivalent recursive representation.              
        }
\end{figure}

We expand the composite metric $g_{\mu\nu}=G_{\mu\nu}+h_{\mu\nu}$ from Eq.~(\ref{eq:gmn}) in powers of $1/(V_0+\Lambda)$, which gives
\begin{equation}
h_{\mu\nu}=\frac{1}{(V_0+\Lambda)}P_{\mu\nu}^{\ \ \alpha\beta}\,:\mathcal{T}_{\alpha\beta}: + \; {\cal O}\left(\frac{1}{V_0+\Lambda}\right)^2,
\end{equation}
 where \begin{equation}
 P_{\mu\nu}^{\ \ \alpha\beta}\equiv \frac{1}{2}\left[(D/2-1)\left(\delta_{\mu}^{\ \alpha}\delta_{\nu}^{\ \beta}+\delta_{\nu}^{\ \alpha}\delta_{\mu}^{\ \beta}\right)-G_{\mu\nu}G^{\alpha\beta}\right].\end{equation}
 With $h_{\mu\nu}\sim\order{1/(V_0+\Lambda)}$, the relevant terms in the action for calculation of the four-point function may be written,
\begin{align} \label{eq:action}
    S &= \int \dd^D x \sqrt{\abs{G}} \bigg( \frac{V_0+\Lambda}{D/2-1} + \frac{1}{2} \sum_{a=1}^N :\partial^{\mu} \phi^a \partial_{\mu} \phi^a: - \sum_{a=1}^N \frac{1}{2}m^2 :\phi^a \phi^a: \nonumber \\ 
    &  - \frac{1}{2}h^{\mu \nu}:\mathcal{T}_{\mu \nu}: + \frac{1}{4}\frac{V_0+\Lambda}{D/2-1}\bigg( h^{\mu \alpha}h^{\nu \beta}-\frac{1}{2}h^{\mu \nu}h^{\alpha \beta}\bigg) G_{\mu \nu} G_{\alpha \beta}\bigg)\bigg]+\order{1/(V_0+\Lambda)}^2\bigg).
\end{align}
To order $1/(V_0+\Lambda)$, collecting terms in the action Eq.~(\ref{eq:action}) then gives a tree-level interaction of the form,
\begin{equation} \label{eq:Lint}
    \mathcal{L}_{\text{int}} = -\frac{1}{4(V_0+\Lambda)}P^{\mu\nu\alpha\beta}\,:\mathcal{T}_{\mu \nu}:\,:\mathcal{T}_{\alpha \beta}:
\end{equation}
In the large-$N$ limit, where $N$ is the number of scalar fields, the leading Feynman diagrams representing the four-point correlation function of $\phi$ fields are displayed in Fig. \ref{fig:pos}. Each interaction vertex includes a factor of $1/(V_0+\Lambda)$, which we recognize from Sec.~\ref{sec:Model} as being of order $1/N$. Quantum gravity corrections are higher-order in $1/N$, as we will discuss in Sec.~\ref{sec:Discussion}.

We write the correlation function to leading order in $1/N$  in the form
\begin{align} \label{eq:corr}
    \bra{0}T\{\phi^a(x_1)\phi^a(x_2) & \phi^c(x_3)\phi^c(x_4) e^{i\int \dd^D x\mathcal{L}_{\text{int}}(x)}\}\ket{0} = \nonumber \\
    & \int \dd^D y_1 \dd^D y_2 \sqrt{\abs{G(y_1)G(y_2)}}\: E_{\mu\nu}(x_1,x_2,y_1) iA^{\mu \nu \alpha \beta}(y_1,y_2)   E_{\alpha\beta}(x_3,x_4,y_2)
\end{align}
where the external factors $E_{\mu\nu}(x_1,x_2,y_1)$ are defined as the connected correlation function
\begin{align} \label{eq:cntr}
 E_{\mu\nu}(x_1,x_2,y_1) \equiv \; & \langle \phi^a(x_1)\phi^a(x_2)\mathcal{T}_{\mu\nu}(y_1)\rangle_{\rm con}.
 \end{align}
The recursion relation represented by Fig. \ref{fig:pos}b takes the form,
\begin{align} \label{eq:recur}
    \int & \dd^D y_1 \dd^D y_2 \sqrt{\abs{G(y_1)G(y_2)}}\:   E_{\mu\nu}(x_1,x_2,y_1)iA^{\mu \nu \alpha \beta}(y_1,y_2)   E_{\alpha\beta}(x_3,x_4,y_2)   =  \nonumber \\
    & \int \dd^D y \sqrt{\abs{G(y)}}\: E_{\mu\nu}(x_1,x_2,y)iA_0^{\mu \nu \alpha \beta}(y)   E_{\alpha\beta}(x_3,x_4,y)  \nonumber \\
    & +    \int \dd^D y_1 \dd^D y_2 \dd^D y_3 \sqrt{\abs{G(y_1)G(y_2)G(y_3)}}\: E_{\mu\nu}(x_1,x_2,y_1) \nonumber \\
    & \qquad \qquad \times \bigg[ {K^{\mu \nu}}_{\lambda \kappa}(y_1,y_2)iA^{\lambda \kappa \alpha \beta}(y_2,y_3)  \bigg]   E_{\alpha\beta}(x_3,x_4,y_3)
\end{align}
The first term on the right-hand-side is the tree-level amplitude, with
\begin{equation} \label{eq:A0}
    {A_0}^{\mu \nu \rho \sigma} = -\frac{1}{4(V_0+\Lambda)}\left[(D/2-1)(G^{\nu \rho}G^{\mu \sigma}+G^{\mu \rho}G^{\nu \sigma}) - G^{\mu \nu}G^{\rho \sigma} \right],
\end{equation}
while the kernel ${K^{\mu \nu}}_{\lambda \kappa}(y_1,y_2)$ corresponds to the portion of the right-hand side in Fig. \ref{fig:pos}b that connects to the shaded blob, and is determined by the connected correlation function of a product of two energy-momentum tensors in the curved-space vacuum:
\begin{equation} \label{eq:kernelTT}
    {K^{\mu \nu}}_{\lambda \kappa}(y_1,y_2) = i{A_0}^{\mu \nu \rho \sigma}(y_1)  \langle \mathcal{T}_{\rho\sigma}(y_1)\mathcal{T}_{\lambda\kappa}(y_2)\rangle_{\rm con}.
\end{equation}
We determine the correlator $\langle \mathcal{T}_{\rho\sigma}(y_1)\,\mathcal{T}_{\lambda\kappa}(y_2)\rangle_{\rm con}$ from the second variation of the effective action:
\begin{equation}
\left(-i\frac{\delta}{\delta h^{\mu\nu}(x)}\right)\left(-i\frac{\delta}{\delta h^{\alpha\beta}(y)}\right)\left[W_{\rm eff}(h^{\rho\sigma}(z))\right] = \frac{\sqrt{G(x)G(y)}}{4}\langle \mathcal{T}_{\mu\nu}(x)\mathcal{T}_{\alpha\beta}(y)\rangle_{\rm con}.
\label{eq:d2Wdh2}
\end{equation}
This gives,
\begin{align} 
 \langle \mathcal{T}_{\mu\nu}(x)\,\mathcal{T}_{\alpha\beta}(y)\rangle_{\rm con} = &
   \bigg[ \frac{1}{(4\pi)^{D/2}} \frac{1}{D-4}\frac{4Nm^{D}}{D(D-2)}\left(G_{\mu\alpha}G_{\nu\beta}+G_{\nu\alpha}G_{\mu\beta}-G_{\mu\nu}G_{\alpha\beta}\right)
     \nonumber \\
   & \  + \frac{1}{(4\pi)^{D/2}}\frac{1}{D-4}
  \frac{4Nm^{D-2}}{3(D-2)} D_{\mu\nu\alpha\beta}
\bigg]\delta^{(D)}(x-y) \nonumber \\ \label{eq:TTsol}
    = \bigg[ \frac{V_0}{(D/2-1)}&\left(G_{\mu\alpha}G_{\nu\beta}+G_{\mu\beta}G_{\nu\alpha}-G_{\mu\nu}G_{\alpha\beta}\right)
   + \frac{4 M_P^{D-2}}{(D-2)} D_{\mu\nu\alpha\beta}(x)\bigg]\, \delta^{(D)}(x-y),
\end{align}

where $M_P$ was defined in Eq.~(\ref{eq:MP}), and  $D_{\mu\nu\alpha\beta}$ is the linearized gravitational wave operator in the curved background with metric $G_{\mu\nu}$, which in $D=4$ has the form \cite{christensen}, \begin{align}
    D_{\mu\nu\alpha\beta}\equiv &\frac12
     \bigg(  G_{\mu \alpha}G_{\nu \beta}\square - \frac{1}{2}G_{\mu \nu} G_{\alpha \beta}\square + R_{\mu \alpha} G_{\nu \beta} + R_{\nu \alpha} G_{\mu \beta} \nonumber \\
    & \qquad \qquad - 2G_{\mu \nu}R_{\alpha \beta} + 2R_{\mu \nu \alpha \beta} - RG_{\mu \alpha}G_{\nu \beta} + \frac{1}{2}R G_{\mu \nu}G_{\alpha \beta} \bigg) .
\end{align}
We then find the kernel from Eq.~(\ref{eq:kernelTT}), \begin{eqnarray}
K^{\mu\nu}_{\ \ \lambda\kappa}(y_1,y_2) &=&\left[i\frac{V_0}{(V_0+\Lambda)}\,\frac12\left(\delta^\mu_{\ \lambda}
\delta^\nu_{\ \kappa}+\delta^\nu_{\ \lambda}
\delta^\mu_{\ \kappa}\right)\right. \nonumber \\
&&\ \ \ +iA_0^{\mu\nu\rho\sigma}(y_1)
  \frac{4M_P^{D-2}}{(D-2)}D_{\rho\sigma\lambda\kappa}(y_1)\bigg]
 \delta^{(D)}(y_1-y_2). 
\end{eqnarray}

Rearranging the recursion relation Eq.~(\ref{eq:recur}), we have
\begin{align} \label{eq:recur2}
    \int \dd^D y & \sqrt{\abs{G(y)}}  \: E_{\mu\nu}(x_1,x_2,y)iA_0^{\mu \nu \alpha \beta}(y)   E_{\alpha\beta}(x_3,x_4,y)    =  \nonumber \\
    & \int \dd^D y_1 \dd^D y_2 \sqrt{\abs{G(y_1)G(y_2)}}\: E_{\mu\nu}(x_1,x_2,y_1)  \bigg[ i\bigg( 1 - \frac{V_{0}}{(V_0+\Lambda)} \bigg)\delta^\mu_\lambda \delta^\nu_\kappa\, \nonumber \\
    & -  \frac{4M_P^{D-2}}{(D-2)}\,i {A_0}^{\mu \nu \rho \sigma} D_{\rho\sigma\lambda\kappa}(y_1)
  \bigg] i A^{\lambda \kappa \alpha \beta}(y_1,y_2)   E_{\alpha\beta}(x_3,x_4,y_2).  
\end{align}
From the explicit form of $A_0^{\mu\nu\rho\sigma}$ we find, \begin{equation}
\frac{(V_0+\Lambda)}{(D/2-1)}\left(G_{\rho\lambda}G_{\sigma\kappa}+G_{\rho\kappa}G_{\sigma\lambda}-G_{\rho\sigma}G_{\kappa\lambda}\right)A_0^{\mu\nu\rho\sigma}=\frac{1}{2}\left(\delta^\mu_{\ \lambda}\delta^\nu_{\ \kappa}+
\delta^\nu_{\ \lambda}\delta^\mu_{\ \kappa}\right). \end{equation}
Using this we can rewrite  Eq.~(\ref{eq:recur2})
 as \begin{align} \label{eq:recur3}
    \int \dd^D y & \sqrt{\abs{G(y)}}  \: E_{\mu\nu}(x_1,x_2,y)iA_0^{\mu \nu \alpha \beta}   E_{\alpha\beta}(x_3,x_4,y)    =  \nonumber \\
    & \int \dd^D y_1 \dd^D y_2 \sqrt{\abs{G(y_1)G(y_2)}}\: E_{\mu\nu}(x_1,x_2,y_1) i {A_0}^{\mu \nu \rho \sigma} \bigg[ \frac{4\Lambda}{(D-2)} \bigg( G_{\rho \lambda}G_{\sigma \kappa} - \frac{1}{2}G_{\rho \sigma}G_{\lambda \kappa}\bigg) \nonumber \\
    & -  \frac{4M_P^{D-2}}{(D-2)} 
    D_{\rho\sigma\lambda\kappa}(y_1)
\bigg]i A^{\lambda \kappa \alpha \beta}(y_1,y_2)   E_{\alpha\beta}(x_3,x_4,y_2).  
\end{align}
The recursion relation will be satisfied if the amplitude $A^{\lambda \kappa \alpha \beta}(y_1,y_2)$ is a solution to the equation,
\begin{align} \label{eq:Green}
   \bigg[\frac{4\Lambda}{(D-2)} \bigg( G_{\rho \lambda}G_{\sigma \kappa} & - \frac{1}{2}G_{\rho \sigma}G_{\lambda \kappa}\bigg) - \frac{4M_P^{D-2}}{(D-2)} D_{\rho\sigma\lambda\kappa}(y_1)
\bigg]i A^{\lambda \kappa \alpha \beta}(y_1,y_2) \nonumber \\
    & \ =  \frac{1}{\sqrt{\abs{G(y_1)}}} \,\frac12\left(\delta^\alpha_\rho\delta^\beta_\sigma +\delta^\beta_\rho\delta^\alpha_\sigma \right)\delta^{(D)}(y_2-y_1).
\end{align}
This is the equation for the Green's function of the  linearized Einstein equations in the vacuum spacetime with cosmological constant $\Lambda$ \cite{christensen}. 
Hence, we have found that the four-point amplitude, summed over the chains of bubbles in Fig.~\ref{fig:pos}, contains the spin-2 propagator of a graviton in the vacuum spacetime with cosmological constant. Note that the gauge-dependent part of the amplitude decouples when attached to the covariantly-conserved energy-momentum tensor from the interaction vertex attached to the external lines. We conclude that a gravitational interaction emerges from local interactions between the $\phi$ bosons. This is our main result.

\section{Discussion}
\label{sec:Discussion}
\subsection{Higher variations of the effective action}
There is one subtlety in the above calculation that we clarify here. We used the relation between the connected correlator $\langle \mathcal{T}_{\mu\nu}(x)\mathcal{T}_{\alpha\beta}(y)\rangle_{\rm con}$ and the second-order terms in the expansion of $W_{\rm eff}$ with respect to $h_{\mu\nu}=g_{\mu\nu}-G_{\mu\nu}$. The subtlety is that the expansion in $h_{\mu\nu}$ does not correspond directly to variations with respect to $g^{\mu\nu}$. 
The second variation of the effective action with respect to $h^{\mu\nu}$  is, \begin{eqnarray}
&&\left(-i\frac{\delta}{\delta h^{\mu\nu}(x)}\right)\left(-i\frac{\delta}{\delta h^{\alpha\beta}(y)}\right)\left[\ln\int {\cal D}\phi\,e^{iS_{\rm free}}\right] \nonumber \\
&=&\int d^Dz\,d^Dw\frac{\delta g^{\rho\sigma}(z)}{\delta h^{\mu\nu}(x)}\left(-i\frac{\delta}{\delta g^{\rho\sigma}(z)}\right)
\left[\frac{\delta g^{\gamma\delta}(w)}{\delta h^{\alpha\beta}(y)}\left(-i\frac{\delta}{\delta g^{\gamma\delta}(w)}\right)\ln
\int {\cal D}\phi\,e^{iS_{\rm free}}\right] \nonumber \\
&=&\frac{\sqrt{g(x)g(y)}}{4}\left(\langle \mathcal{T}_{\mu\nu}(x)\mathcal{T}_{\alpha\beta}(y)\rangle-\langle \mathcal{T}_{\mu\nu}(x)\rangle\langle \mathcal{T}_{\alpha\beta}(y)\rangle\right) \nonumber \\
&&-i\frac{\int d^Dz\,d^Dw\,\frac{\delta g^{\rho\sigma}(z)}{\delta h^{\mu\nu}(x)}\frac{\delta g^{\gamma\delta}(z)}{\delta h^{\alpha\beta}(x)}\,\int{\cal D}\phi\,e^{iS_{\rm free}}\frac{1}{2}\frac{\delta}{\delta g^{\rho\sigma}(z)}\left(\sqrt{g(w)}T_{\gamma\delta}(w)\right)}{\int{\cal D}\phi\,e^{iS_{\rm free}}} \nonumber \\
&&-i\int d^Dw \,\left\langle\frac{\sqrt{g(w)}}{2}\mathcal{T}_{\gamma\delta}(w)\right\rangle
\,\frac{\delta^2 g^{\gamma\delta}(w)}{\delta h^{\mu\nu}(x)\,\delta h^{\alpha\beta}(y)}
. \label{eq:d2W}
\end{eqnarray}
In the last term of Eq.~(\ref{eq:d2W}) we use, \begin{equation}
g^{\mu\nu}=G^{\mu\nu}-h^{\mu\nu}+h^{\mu\alpha}h_\alpha^{\ \nu} +{\cal O}(h^3), \end{equation}
where indices are raised and lowered with $G_{\mu\nu}$.
In the next-to-last term of Eq.~(\ref{eq:d2W}) we use Eq.~(\ref{eq:<dphidphi>}) to obtain, \begin{equation}
\frac{\delta}{\delta g^{\rho\sigma}(z)}\left\langle\sqrt{g(w)}\mathcal{T}_{\gamma\delta}(w)\right\rangle=-\frac{1}{2}\delta^{(D)}(z-w)\sqrt{g(w)}\langle g_{\rho\sigma}(w)\mathcal{T}_{\gamma\delta}(w)+g_{\gamma\delta}(w)\mathcal{T}_{\rho\sigma}(w)\rangle.
\end{equation}
Then it is straightforward to see that the last two terms of Eq.~(\ref{eq:d2W}) cancel, leaving, \begin{equation}
\left(-i\frac{\delta}{\delta h^{\mu\nu}(x)}\right)\left(-i\frac{\delta}{\delta h^{\alpha\beta}(y)}\right)\left[\ln\int {\cal D}\phi\,e^{iS_{\rm free}}\right] = \frac{\sqrt{g(x)g(y)}}{4}\langle \mathcal{T}_{\mu\nu}(x)\mathcal{T}_{\alpha\beta}(y)\rangle_{\rm con}.
\end{equation}

\subsection{Quantum gravity corrections}
\begin{figure}[ht] 
        \includegraphics[scale=0.3,trim=0 270 0 100, clip]{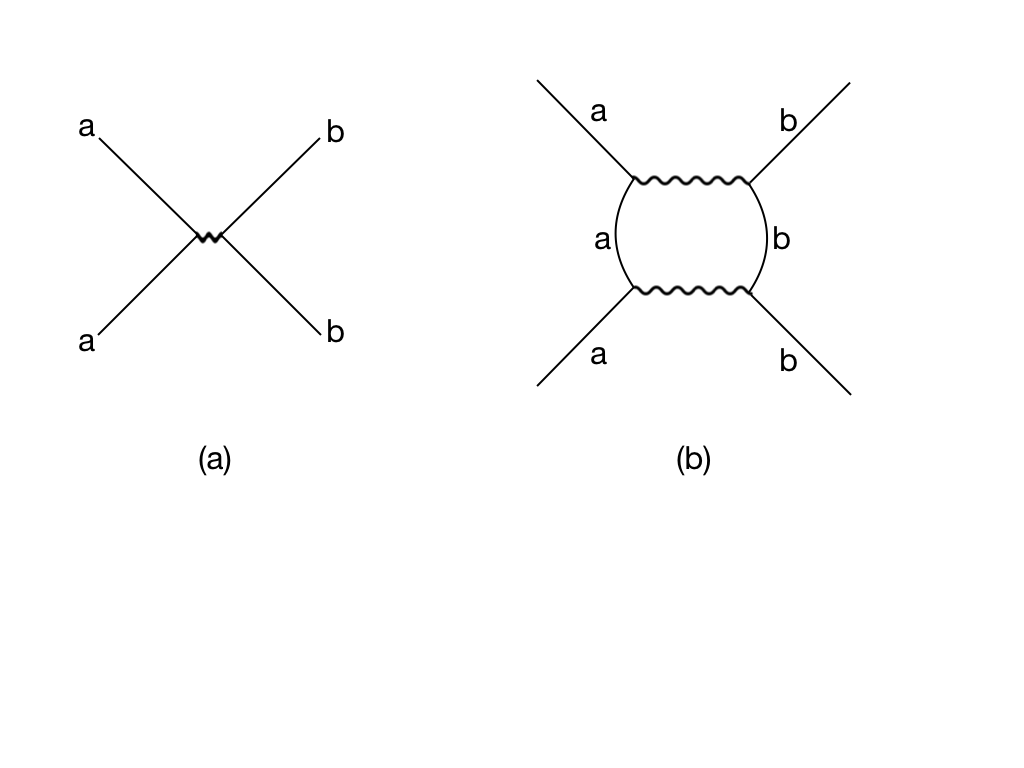}
        \caption{
                \label{fig:interaction} 
                Quantum gravity corrections. (a) The wiggly line at the $T_{\mu\nu}-T_{\alpha\beta}$ vertex indicates that any insertion of the vertex comes together with the sum of loops that gives rise to the graviton propagator. (b) Diagrams of higher-order in $1/N$ are quantum gravity corrections involving composite gravitons in loops.              
        }
\end{figure}

We organized the calculation by way of an expansion in $1/N$ in order to identify the emergent gravitational interaction, but $N$ need not be large for the existence of the spin-2 graviton state. We can identify  quantum gravity corrections as $1/N$ corrections. For this purpose, we draw the interaction vertex as in Fig.~\ref{fig:interaction}a. Factors of $N$ come from loops and from the vertex, which includes a factor of $1/(V_0+\Lambda)\sim \mathcal{O}(1/N)$. Any insertion of the interaction vertex in a Feynman diagram comes with additional diagrams in which the wiggly vertex is replaced by a chain of loops as in Fig.~\ref{fig:pos}a. Then, for example, the $1/N$-suppressed contribution to the four-point function 
in Fig.~\ref{fig:interaction}b we recognize as belonging to a two-graviton box correction to the correlation function. Quantum gravity corrections are suppressed at large $N$, as also follows from the fact that the derived Planck mass $M_P$ in Eq.~(\ref{eq:MP}) is proportional to $N$.

\subsection{A comment on the number of degrees of freedom}
The theory we have analyzed is defined as a diffeomorphism-invariant theory of $N$ scalar fields $\phi^a$. In $D$ dimensions, it would seem that $D$ of these degrees of freedom should be nondynamical, fixed by a gauge-fixing of the coordinates. However, in the present approach, $D$ clock and rod fields $X^M$ were introduced in Eq.~(\ref{eq:gmn}) in order to fix a coordinate basis in which the vaccum spacetime metric is specified, and the composite metric is gauge-fixed when defined in terms of the fundamental fields. All of the fields $\phi^a$ are dynamical in this theory. 

The clock and rod fields are indeed nondynamical, but in a stricter sense than being constrained by gauge-fixing: The clock and rod fields do not contribute to the action defining the theory. They are not just nondynamical,  they are unphysical. Instead, a term involving the clock and rod fields was added to and then subtracted from the composite metric appearing in the action in order to enable a perturbative expansion about the curved-space vacuum. Subtracting off the contribution of the clock and rod fields also served to cancel a tadpole involving derivatives of the physical fields $\phi^a$ in the composite metric, in analogy with a related analysis in Ref.~\cite{Carone:2019xot}. The cancellation of this divergent tadpole was necessary for consistency of the perturbative expansion in this approach. In this theory, restoration of diffeomorphism invariance by decoupling the clock and rod fields is tantamount to normal ordering of derivative terms in the composite metric.

A curious corollary of this observation regards the same theory with $N=D$ scalar fields. On the one hand, the theory is topological in the sense that the only diffeomorphism-invariant information in the field configuration is the number of critical points at which $\det\left(\partial \phi^a/\partial x^\mu\right)=0$ and the value of the fields at those points. On the other hand, expanding about the vacuum-expectation-value of the composite metric defined in terms of those fields, the above analysis demonstrates that the regularized topological theory contains a composite graviton state and an emergent gravitational interaction, albeit with potentially large quantum gravity corrections, as this theory is far from the large-$N$ limit.

\subsection{Generalization to other theories}
Here we analyzed a scalar toy model, which allowed us to elucidate the dynamical selection of the vacuum spacetime and the emergent gravitational interaction in this framework. However, presuming a physical ultraviolet regulator, the generalization to theories whose low-energy description is that of a given field theory coupled to gravity has been suggested before \cite{Carone:2016tup,Erlich:2018qfc}. To summarize the algorithm, with the benefit of our present understanding of the decoupling of the clock and rod fields from the dynamics: Beginning with a Lorentz-invariant quantum field theory, \begin{itemize}
    \item Minimally couple the theory to a spacetime specified by a metric $g_{\mu\nu}$, or vielbein $e_{\mu}^{a}$ in a theory with fermions.
    \item Identify the composite metric $g_{\mu\nu}$, or vielbein $e_{\mu}^a$ up to local Lorentz transformations, in terms of the fundamental fields by solving the constraint $T_{\mu\nu}=0$. In general this cannot be done analytically.
    \item Add and subtract the background vielbein $E_{\mu}^a=\left(\partial_\mu X^M\right) E_{M}^a$ to the composite vielbein operator in analogy with Eq.~(\ref{eq:gmn}) in order to organize a perturbative expansion about the vacuum.
    \item Determine the vacuum spacetime self-consistently by imposing $\langle T_{\mu\nu}[E_{\mu}^a,\Phi]\rangle=0$ in the vacuum, where $\Phi$ represents the fundamental fields in the theory.
\end{itemize}
This procedure gives rise to a diffeomorphism-invariant theory which, when expanded about the vacuum, includes the original Lorentz-invariant theory in the low-energy effective description. Due to  diffeomorphism invariance, we expect the theory so defined to contain an emergent gravitational interaction in the self-consistent vacuum. In a particular generalization with scalars and fermions, with effective cosmological constant $\Lambda$ tuned to zero, the emergent gravitational interaction was analyzed in  an analogous way in Ref.~\cite{Carone:2018ynf}.

\section{Conclusions}

We have demonstrated that a diffeomorphism-invariant scalar theory with an ultraviolet regulator has a self-consistent vacuum spacetime determined by Einstein's equations with higher-derivative corrections, and an emergent gravitational interaction in the background of the vacuum spacetime. The construction is diffeomorphism invariant, and nonlinear gravitational self-interactions arise as in Ref.~\cite{Carone:2017mdw}. It should be straightforward to generalize this analysis to an arbitrary theory with fermions and gauge fields. Beginning with an arbitrary field theory in curved spacetime, the composite vielbein in this approach is determined (up to local Lorentz transformations) as a function of the fields by the condition $T_{\mu\nu}[e_\alpha^{\ a},\Phi]=0$ as an operator equation, where $e_\alpha^{\ a}$ is the composite vielbein and $\Phi$ represents the fundamental fields in the theory. 

We introduced {\em nondynamical} clock and rod fields, which provide a coordinate basis with which to describe the vacuum spacetime. 
Adding terms with derivatives of the clock and rod fields to the action, with field-space metric of the same form as the vacuum spacetime metric, permits a perturbative expansion about the vacuum. However, subtracting those same terms from the action then precisely cancels a tadpole that would otherwise contribute to the vacuum expectation value of the energy-momentum tensor, in conflict with a basic principle of this approach to quantum gravity.

For ease of discussion we dropped  higher-derivative corrections in the effective action and correlation functions. Including those corrections would modify both Einstein's equations for the vacuum spacetime, and the kernel of the recursion relation that we used to determine the four-point correlation function. With the ultraviolet regulator held fixed, the corrections to the low-energy effective description are important at curvatures large compared to $m^2$. The analogy of these corrections to a more realistic scenario that incorporates the standard model particle content would be relevant for early cosmology and other circumstances involving strong gravitational effects.

To serve as a complete quantum theory of gravity, this approach requires a physical regulator rather than dimensional regularization, point splitting, or the like. In an earlier work it was suggested that a  stochastic evolution of fields might provide the fundamental description of a theory of this type, and the discreteness of the stochastic process would provide the ultraviolet regulator \cite{Erlich:2018qfc}. A precise formulation and simulations of such a theory would shed light on the dynamical evolution of generic states relative to the fiducial clock and rod fields. Note that the existence of a tentative physical regulator does not automatically imply a smooth semiclassical description of the theory, a difficult lesson that was learned in the context of the dynamical triangulations approach to quantum gravity \cite{Loll:1998aj,Ambjorn:2004qm}.

Finally, we note that if gravitation is an emergent interaction not present at short distances, then the equilibrium configuration of an initially dense state may be homogeneous and isotropic, rather than clumped as it would be under the influence of gravitation. If we erroneously assume that gravitation exists as a fundamental interaction at short distances then a homogeneous, isotropic state would appear to have an anomalously small entropy \cite{Penrose}. Hence, composite gravity, and emergent gravity scenarios more generally, may explain the past hypothesis by replacing the requirement of a low-entropy initial state with one of high density or temperature. Possible implications of a gravitational phase transition for early-universe cosmology would be interesting to explore.

\label{sec:Conclusions}

\begin{acknowledgments}  
This work was supported by the NSF under Grant PHY-1819575. We are grateful to Chris Carone and Diana Vaman for many useful conversations, and to Chris Carone for comments on the manuscript.
\end{acknowledgments}

\end{document}